\documentclass{emulateapj}
\usepackage{graphics,graphicx}
\def\kms{km\,s$^{-1}$}
\def\etal{ et~al.\rm}

\def\mjysr{MJy sr$^{-1}$}

\def\g292{G292.0+1.8}
\def\spitzer{{\it Spitzer}}
\def\ergss{ergs s$^{-1}$}
\def\chan{{\it Chandra}}

\tighten
\begin{document}

\title{\bf \spitzer\, Imaging and Spectral Mapping of the Oxygen-Rich Supernova Remnant G292.0+1.8  }

\author{Parviz Ghavamian\altaffilmark{1}, Knox S. Long \altaffilmark{2},  William P. Blair\altaffilmark{3}, Sangwook Park\altaffilmark{4},
Robert Fesen\altaffilmark{5}, B. M. Gaensler\altaffilmark{6},  John P. Hughes\altaffilmark{7}, Jeonghee Rho\altaffilmark{8},
and P. Frank Winkler\altaffilmark{9} 
  }

\accepted{February 23, 2012}

\altaffiltext{1}{Department of Physics, Astronomy and Geosciences, Towson  University, Towson, MD, 21252}
\altaffiltext{2}{Space Telescope Science Institute, 3700 San Martin Drive, Baltimore, MD, 21218}
\altaffiltext{3}{Department of Physics and Astronomy, Johns Hopkins University, 3400 N. Charles Street, Baltimore, MD, 21218}
\altaffiltext{4}{Department of Physics, Box 19059, University of Texas at Arlington, Arlington, TX, 76019}
\altaffiltext{5}{Department of Physics and Astronomy, Dartmouth College, 6127 Wilder Lab, Hanover, NH 037355}
\altaffiltext{6}{Sydney Institute for Astronomy, School of Physics A29, The University of Sydney, NSW 2006, Australia }
\altaffiltext{7}{Department of Physics and Astronomy, Rutgers University, 136 Frelinghuysen Road, Piscataway, NJ, 08854}
\altaffiltext{8}{NASA Ames Research Center, Moffett Field, CA, 94035 }
\altaffiltext{9}{Department of Physics, Middlebury College, McCardell Bicentennial Hall 526, Middlebury, VT, 05753}

\begin{abstract}

We present mid-infrared continuum and emission line images of the Galactic oxygen-rich supernova remnant (SNR) \g292, 
acquired using the MIPS and IRS instruments on  the \spitzer\, {\it Space Telescope}.  The MIPS 24 $\micron$ and 70 $\micron$ images of \g292\,
are dominated by continuum emission from a network of filaments encircling the SNR.  The morphology of the SNR, as seen in the mid-infrared, 
resembles that seen in X-rays with the \chan\, X-ray Observatory.   Most of the mid-infrared emission in the MIPS images is produced by 
circumstellar dust heated in the non-radiative shocks around \g292, confirming the results of earlier mid-IR observations with AKARI.   
In addition to emission from hot dust, we have also mapped 
atomic line emission between 14 $\micron$ and 36 $\micron$ using IRS spectral maps.  
The line emission is primarily associated with the bright oxygen-rich optical knots, but is also detected 
from fast-moving knots of ejecta.  We confirm our earlier detection of 15-25 $\micron$ emission characteristic of magnesium silicate dust
in spectra of the radiatively shocked ejecta.  
We do not detect silicon line emission from any of the radiatively shocked ejecta
in the southeast of the SNR, possibly because that the reverse shock has not yet penetrated most of the Si-rich
ejecta in that region.  This may indicate that \g292\, is less
evolved in the southeast than the rest of the SNR, and may be further evidence in favor of an asymmetric SN explosion as proposed
in recent X-ray studies of \g292.

\end{abstract}


\section{INTRODUCTION}
Oxygen-rich supernova remnants (SNRs) are objects whose optical spectra are dominated by oxygen forbidden line emission
(i.e., [O~I], [O~II] and [O~III]).  This emission arises in radiative shocks in the oxygen-rich ejecta expelled from the core-collapse supernova (SN) explosion.  
\g292, which is  the result of an 
unrecorded SN that took place about 3000 years ago (Ghavamian, Hughes \& Williams 2005; Winkler \& Long 2006), is one of only seven such O-rich SNRs
known today.   In [O~III] $\lambda$5007 optical images \g292\, is dominated by a distinct, crescent-shaped structure approximately 1\arcmin\, in size
(hereafter the `Spur') located on the eastern side of the SNR (Goss \etal\, 1979).  A collection of localized clumps 
(fast-moving knots, or FMKs, similar to those seen in Cas A) have also been found in the interior of \g292\,
(Ghavamian, Hughes \& Williams 2005; Winkler \& Long 2006).  This contrasts with the
appearance of the SNR at X-ray (Clark, Tuohy \& Becker 1980; Park \etal\, 2002; 2004, 2007; Gonzalez \& Safi-Harb 2003) and radio (Gaensler \& Wallace 2003) wavelengths, where the remnant appears as a slightly elliptical shell approximately
8\arcmin\, across.   

The X-ray emission from O-rich SNRs, including \g292, tends to arise from 
faster, non-radiative shocks in lower density ejecta and interstellar gas.   \g292 has a complex X-ray morphology, with wide-spread 
shocked circumstellar material superposed on a network of shocked ejecta knots.   The SNR also features a filament (or filaments) stretching 
east-west across the middle of the SNR (commonly called the `equatorial belt' in earlier papers).  
Analyses of \chan\ data indicate the equatorial belt is of normal
composition, suggesting shocks propagating in circumstellar material (Park \etal\, 2002, 2004; Gonzalez \& Safi-Harb 2003; Lee \etal\, 2010).  

\g292 also hosts a pulsar (PSR J1156-5916) with a spin down age of 2900 years (Camilo \etal\, 2002). The pulsar has been detected over much 
of the electromagnetic spectrum (see Camilo \etal\, 2002 for radio observations, Hughes \etal\, 2003 for X-rays), including  
4.5 and 8.0 $\micron$ with the IRAC imager on \spitzer\ (Zyuzin \etal\, 2009).   The pulsar wind nebula of PSR J1156-5916 has
also been detected in the radio (Gaensler \& Wallace 2003), IR (IRAC imagery at 4.5 $\micron$ and 8.0 $\micron$; Zyuzin \etal\, 2009) and
in the X-rays (Hughes \etal\, 2001).  With all these properties and as the second youngest
O-rich SNR known in the Galaxy, \g292 is an important object for understanding how core-collapse SNRs evolve.

Here we report on a set of imaging observations of \g292\ obtained with the MIPS and IRS instruments on \spitzer.  The MIPS 24 $\micron$ and
70 $\micron$ images trace mostly emission from warm dust heated by shocks in the shocked circumstellar medium (CSM).  Our observations also included
IRAC imagery at 4.5 and 8.0 $\micron$, though no discernible emission from \g292\, was detected in these two wavebands (consistent with non-detection
at similar wavelengths in the AKARI observations (Lee \etal\, 2009).  Narrow band images 
constructed from our IRS spectral maps trace emission from the strongest mid-IR emission lines and provide clues to the location of ejecta within
the SNR.   

The observations reported here are part of a follow-up to our earlier IRS observations of two positions in \g292  
(Ghavamian et al.\ 2009).  The spectra obtained earlier showed emission lines from [Ne~II] $\lambda$12.8, [Ne~III] $\lambda\lambda$15.5,36.0,
[Ne~V] $\lambda$24.3, and [O~IV] $\lambda$25.9, but no clear evidence of emission from heavier elements.  This contrasts with Cas A, where in
addition to these lines, IRS 
spectra showed significant [Ar~II] $\lambda$7.99, [Fe~II] $\lambda$17.9 and [S~III] $\lambda\lambda$18.7, 33.5 emission (Rho \etal\, 2008;
Smith \etal\, 2009).   
Our goal here is to provide a more general description of the SNR as a whole in the mid-IR.  

The \spitzer\, observations in this paper complement the existing AKARI study
of \g292\, by Lee \etal\, (2009) where mid-IR imagery was acquired of the SNR in 10 bands centered at wavelengths ranging
from 2.7 $\micron$ to 180 $\micron$ (including 24 $\micron$ band with similar sensitivity and spatial resolution to the MIPS
24 $\micron$ images presented here).  In addition to the imagery, AKARI spectra were obtained from a section of the belt exhibiting
the brighest X-ray emission, as well as the lower portion of the O-rich Spur seen in the [O~III] imagery of \g292.  Our
new IRS spectral maps cover the entire SNR over the 5 $\micron$ - 36 $\micron$ range, providing 
access to emission lines not covered in the AKARI observations such as
[O~IV]+[Fe~II] $\lambda$25.9, [Ne~III] $\lambda$15.5 $\micron$ (which falls in a gap in AKARI spectral coverage between 13 $\micron$
and 18 $\micron$) and [Si~II] $\lambda$34.8.  In addition, unlike Lee \etal\, (2009) who focused on properties of the 
integrated IR emission from the entire SNR, our analysis includes analysis of shocked CSM in localized
regions within \g292.

The remainder of this paper is organized as follows:  In Section 2, we describe the observations and the techniques used to reduce the data, 
including the creation of narrow band images from the IRS data.  In Section 3, we describe the broad band images and compare them to X-ray and optical 
images of the SNR.   In Section 4, we discuss the emission
line images, and discuss these primarily in the context of ejecta from the SNR.

\section{OBSERVATIONS AND REDUCTIONS}

Observations of \g292\, described here were performed during Cycle 4 of \spitzer\, (PID 40583; P. Ghavamian, PI) and utilized the 
MIPS (24 $\micron$ and 70 $\micron$) and IRS instruments
(Long-Low module only) in mapping mode.  The MIPS data were obtained on 2008 March 13 (70 $\micron$) and 2008 April 15 (24 $\micron$),
while the IRS data were taken on 2008 August 13.
The MIPS raster maps at 24 $\micron$ and 70 $\micron$ covered the entire SNR (8\farcm3 across, or
14 pc at an assumed distance of 6 kpc, Gaensler \& Wallace 2003), as well as a sizeable swath 
of the surrounding sky.  The MIPS 24 $\micron$ observations were performed in a one-cycle raster
map, with  an exposure time of 10 s per pixel and a total integration time of 500 s.  The 70 $\micron$
observations were also obtained in a one-cycle raster map, with an exposure time of 10s  per pixel and
total integration time of 380s.  The 1-$\sigma$ extended source sensitivity of the MIPS observations was approximately 0.2 \mjysr\, at 24 $\micron$
and 4 \mjysr\, at 70 $\micron$.  The MIPS images of \g292\, are shown in Figure~1.

The IRS spectral maps utilized
one cycle of 5 pointings parallel to the LL slit, separated by 120$\arcsec$, along with 112 pointings taken in 6$\arcsec$
steps perpendicular to the slit.  Both LL1 (1st order, 19.5-38.0 $\micron$) and LL2 (2nd order, 14.0-21.3 $\micron$) were active
during the mapping scans, with an exposure time of 32s per 
pixel (560 individual spectra).   The IRS mapping footprint for these observations is shown overlaid onto the MIPS 24 $\micron$ image
in Figure~2.

\subsection{Post-Processing of MIPS data}

Our MIPS data were processed using calibration pipeline version S18.12.0.  MIPS delivers diffraction limited images, so
that the relative spatial resolution  of any two MIPS channels differs by the ratio of their central wavelengths.  To compare the surface brightnesses of features 
between images, we first degraded the spatial resolution of the 24 $\micron$ image to that of the 70 $\micron$ image
by convolving the former with a PSF kernel using the IDL-based Convolution Kernels software CONVIMAGE (Gordon \etal\, 2008).  We
used a PSF kernel appropriate for a 50~K blackbody source.   While we do not expect the continuum emission to follow a simple
blackbody shape, this temperature is approximately midway between the temperatures of the cold ($\sim$30~K)
and warm ($\sim$70~K) CSM dust components fit to the IRS staring mode spectra of \g292\, by Ghavamian \etal\, (2009).  After the convolution we
used the AIPS (Astronomical Image Processing System)\footnote{AIPS is produced and supported by the National Radio Astronomy Observatory, operated by Associated Universities, Inc., under
contract with the National Science Foundation} task HGEOM to resample the convolved 24 $\micron$ image
(2\farcs55 pixel$^{-1}$) onto a grid matching that of the 70 $\micron$ image (5\farcs3 pixel$^{-1}$).  
To extract surface brightness values from selected regions around \g292\, we utilized the FUNTOOLS package of SAO 
({\sf https://www.cfa.harvard.edu/\textasciitilde john/funtools/}).

To compare localized variations in the 70/24 flux ratios with the corresponding X-ray emission in \g292, we estimated the X-ray
brightnesses using the deep 510 ks \chan\, image of \g292\, (Park \etal\, 2007).   We started with the \chan\, level 1 event file from 
that observation, processed in the manner described in Park \etal\, (2007) and filtered over the 0.3-8.0 keV range.  Before extracting the X-ray fluxes we
blurred the filtered \chan\, image to the same resolution as the MIPS 24 $\micron$ image.  The pixel scale of
IRAC images is 0\farcs6 pixel$^{-1}$, which is close to the 0\farcs5 pixel$^{-1}$ pixel scale of the \chan\, images.  Therefore,
we convolved the X-ray image using the CONVIMAGE kernels appropriate for blurring an IRAC 3.6 $\micron$ image to MIPS
24 $\micron$ resolution.   We again used the convolution kernel appropriate for a 50~K blackbody.
Finally, we extracted the X-ray counts from individual regions (18 regions were selected, described later in Section 4) in the convolved
X-ray image using the FUNCNTS application.  We utilized
an annular region surrounding \g292\, in the \chan\, image to estimate the underlying background level (marked in Figure~3).  
Although the background region includes a number of faint point sources, their total contribution to the background counts is negligible
($\lesssim$2\%), so their contribution was not removed before scaling and subtracting the background from each extraction region.
After background subtraction, we converted the resulting net counts to count rates
by dividing by the exposure time (510 ks).  We then converted the count rates to surface brightness using the PIMMS tool from CXC.
Note that the X-ray fluxes
are estimated without actual spectral modeling, resulting in less accurate flux estimates than would be obtained with spectral models.
However, we only seek crude flux estimates for the purpose of identifying systematic trends in the ratios of IR to X-ray emission.
Since the fluxes are extracted from CSM shocks, we assumed
abundances 0.2 times solar, while taking $N_H\,=\,$6$\times$10$^{21}$ cm$^{-2}$ and $k T\,=\,$0.75 keV.  These parameters
reflect average values for these parameters measured around the rim of \g292\, by Lee \etal\, (2010).

\subsection{Post-Processing of IRS data}

We performed our IRS spectral mapping analysis using data processed with \spitzer\, calibration pipeline version S18.7.0.
Using the IDL CUBISM software (Smith \etal\, 2007), we assembled the 560 Basic Calibrated Data files (BCDs)
into two data cubes, one for LL1 and one for LL2.  The BCDs are
heavily affected by hot pixels (especially beyond 35 $\micron$), which results in vertical columns of elevated pixels in the
assembled data cube.  To mitigate the effects of these hot pixels, we median smoothed the BCDs prior to creating the data cubes
by using the $FILTER\_IMAGE$ routine
from the IDL Astronomy User's library.  The smoothing replaces the value of each pixel with the median of the surrounding
3 pixels in a moving box.  While this reduces the spatial resolution of the data in the BCDs somewhat, it 
significantly mitigates the hot pixel columns in the resulting datacube.

The background emission around \g292\, varies noticeably
across the face of the SNR in both MIPS images.  At 24 $\micron$, the background is about 17.6 \mjysr\,  in the northeastern corner of the image in the
region used for estimating the background emission in the data cubes.  The background rises to a maximum of 18 \mjysr\, in the
southwestern corner of the image.  The corresponding background surface brightnesses in the 70 $\micron$ image are 32 \mjysr\, and 35 \mjysr,
respectively. This emission gradient can be seen in 
Figure~1.  The variability is due in part to the presence of an H~II region along the line of
sight just south of \g292.  It appears as patchy diffuse emission in the narrowband H$\alpha$, [S~II] and [O~III]
images of the SNR (Winkler \& Long 2006).   

In the IRS datacubes the H~II region along the line of sight to \g292\, contributes [Ne~III] $\lambda\lambda$15.5, 35.0,
[S~III] $\lambda\lambda$18.7, 35.0 and [Si~II] $\lambda$34.8 line emission, as well as a photo-excited dust continuum 
starting near 15 $\micron$ and slowly rising beyond the edge of the IRS bandpass at 40 $\micron$.
The sky spectrum also shows a cluster of faint, closely spaced emission features between 15 $\micron$ and 20 $\micron$.
These features are characteristic of PAH emission from the intervening photodissociation regions at 16.4 and 17.0 $\micron$,
as well as H$_2$ S(1) 17.1 $\micron$ emission.  At longer wavelengths, faint H$_2$ 28.2 $\micron$ emission is also detected
in the sky spectrum.  The excess contribution from this diffuse component, after subtraction of background from the
northeastern corner of the datacube, results in residual emission in the datacube when 
extracting emission line images.

\section{ANALYSIS}

\subsection{MIPS Imagery}

In Figure~1 we present the 24 $\micron$ and 70 $\micron$ MIPS images of \g292.  The SNR appears
in both bands as an elliptical shell with a banded structure running E-W across its middle, similar to what is observed in
the \chan\, images (Park \etal\, 2002; 2004; 2007; Lee \etal\, 2010).   For the most part, the morphology and brightness 
variations of the shell 
match those seen in the X-rays, with the main difference being the lack of prominent IR emission from X-ray
emitting O-rich and Ne-rich ejecta.  These results are consistent with continuum emission from circumstellar dust heated by the
non-radiative forward shock in \g292.  Lee \etal\, (2009) reached the same conclusion based on their AKARI imagery of
\g292.  

Both the Spur and the FMKs can be seen in the 24 $\micron$ image, tracing the [O IV] $\lambda$25.9 produced by radiative ejecta shocks.  
The [O~IV] emission from these shocks closely follows the distribution of [O~III] $\lambda$5007 emission 
observed in narrowband optical imagery (Tuohy, Burton \& Clark 1982; Winkler \&
Long 2006) and optical imaging spectrometry (Ghavamian, Hughes \& Williams 2005).   
In addition, both the Spur (and as we show in Section 3.2.2, the southernmost FMK) exhibit 
emission from a spectral bump between 15 and 25 $\micron$ in our IRS spectral maps.  
This bump (likely a signature of ejecta dust heated in the radiative shocks) also contributes emission
to the MIPS 24 $\micron$ image of \g292.

The flux ratio between IR images at two different wavelengths is sensitive to such parameters as the dust temperature, gas density and dust-to-gas
ratio (Dwek 1987). To investigate the global 70/24 flux ratio in \g292,
we used the FUNCNTS application from FUNTOOLS to integrate the 24 $\micron$ and 70 $\micron$ emission across the
face of \g292\, (avoiding bright stars within the remnant periphery).  We then estimated the total flux and luminosity
of the SNR in these bands.
We subtracted the sky contribution in each band using an annular region encircling \g292\, (marked in Figure~3; the same background annulus
was used for both the 24 $\micron$ and 70 $\micron$ images).  Similar to the X-ray data,
the background annuli in our 24 $\micron$ images contain point sources (none are present in the 70 $\micron$ images).  These
point sources contribute $\lesssim$3\% to the total counts in the backround regions.  After scaling and subtracting the background,
the resulting fluxes are 9.8 Jy at 24 $\micron$ and
26.4 Jy at 70 $\micron$, respectively.  The globally averaged 70/24 ratio is 2.7, an intermediate value between the local minimum 
of 1.4 and local maximum of 5.5 reported in Table~1.  These fluxes are in good agreement with the two-temperature modified blackbody
SED (a mixture of graphite and silicon) calculated by Lee \etal\ (2009)
for a 3000 year old SNR with preshock density of 0.5 cm$^{-3}$.  Assuming a distance of 6 kpc to \g292\, (Gaensler \& Wallace 2003), the
corresponding luminosities in the MIPS bands are $L_{24}\,=\,$1.9$\times$10$^{36}$ ergs s$^{-1}$ and $L_{70}\,=\,$
1.4$\times$10$^{36}$ ergs s$^{-1}$ (Note the luminosity at 24 $\micron$ includes the contribution of [O~IV]+[Fe~II]
line emission near 25.9 $\micron$).

Localized variations in the 70/24 flux ratio can be observed both along the shell and throughout the interior of \g292.
The 24 $\micron$ and 70 $\micron$ surface brightnesses are shown in Table~1 for a selection of individual regions,
along with the corresponding 70/24 ratios.  A trend seen in these ratios is
that they tend to be largest (i.e., implying coldest dust) along the elliptical outer blast wave of \g292.
The hottest CSM dust tends to be found along the equatorial belt, though some
of the clumpy belt material also extends toward the southwest of the SNR.

A conspicuous difference between
the IR and X-ray appearance of \g292\, is the presence of strong IR emission in parts of the shell lacking
X-ray emission.  These differences can be seen in Figure~3, where the individual regions
have been marked on the MIPS 24 $\micron$ and \chan\, 0.3-8.0 keV images.  Regions 6-10 in the southwestern portion of \g292\, mark
clumps where the IR emission in both MIPS images is particularly strong compared to the X-ray emission.
The spectral properties of these clumps are clearly different from those of the circumstellar belt (numbered 13-18 in Figure~3)
and most of the shell (Regions 1-4 and 11-12).
In contrast, sections of the circumstellar belt (Regions 13-18) are prominent
both in the MIPS and \chan\, images.  These suggest significant differences between physical conditions in the
southwest and those in the rest of the shell and circumstellar belt.
X-ray spectra of Regions 13-18 extracted from the 510 ks \chan\,
observation (Park \etal\, 2007) exhibit no evidence of ejecta (enhanced metal) abundances, and in fact appear fully consistent with
cosmic (subsolar) abundances.
In addition, the southwestern clumps are not detected in either the [O~III] image of \g292\ (Winkler \& Long 2006), 
nor in [O~IV]+[Fe~II], [Ne~III] or [Si~II] in our IRS maps of \g292, an indication their emission arises in non-radiative shocks.

\subsection{Narrowband Maps}

\subsubsection{Emission Line Images}

In addition to the IR continuum generated by the shock heated dust in \g292, the SNR is also detected
in IR forbidden line emission from shocked ejecta.  The detected lines were described
by Ghavamian \etal\, (2009), who reported strong O and Ne emission (as well as possible weak S line emission)
from the Spur.  Clear variations can be seen in the relative fluxes of the ejecta lines in the IRS spectral maps, reflecting
variations in physical conditions in the ejecta.  To map these variations, we first
used CUBISM to extract emission line images from the IRS spectral map.  

Due to the presence of underlying
dust continuum throughout most of the LL2 and all of the LL1 bandpass, isolating the emission line component
required estimation of the continuum level under each line, then the
subtraction of this continuum at each position in the datacube.  To perform this subtraction, we first used CUBISM to isolate
the continuum emission over a narrow sub-band on either side of the [Ne~III] $\lambda$15.6 and [O~IV]+[Fe~II] $\lambda$25.9 lines.
The corresponding sub-bands used for the [Si~II] image were both chosen from continuum on the blue
side of 34.8 $\micron$ to minimize the impact of hot pixels on the continuum-subtracted [Si~II] image.

Collapsing the emission in each of the two bands, we generated
two `off-band' images of \g292\, near each spectral line.  We then averaged the two images to approximate
an image of the continuum underlying each of the emission lines.  We scaled and subtracted this averaged
image from that formed by integrating emission from the \g292\, datacube over each of the O, Ne and Si lines.  
The scaling factor applied to each background image before subtraction was (1.05,1.0,1.10) for [Ne~III], [O~IV]+[Fe~II], and 
[Si~II] respectively.  The wavelength ranges integrated for estimating the underlying [Ne~III] $\lambda$15.5 continuum in the datacube 
were 15.0-15.2 $\micron$ and 15.9-16.1 $\micron$, while that for [O~IV]+[Fe~II] $\lambda$25.9 were 23.4-23.9 $\micron$
and 26.9-27.4 $\micron$.  The corresponding ranges for [Si~II] $\lambda$34.8 $\micron$ were 31.4-32.5 $\micron$
and 32.0-32.7 $\micron$.  The [Ne~V] $\lambda$24.3 emission detected in 
\g292\, (Ghavamian \etal\, 2009) was not strong enough to allow creation of useful maps once the nearby
continuum was subtracted.  

Despite the presence of strong [S~III] $\lambda\lambda$18.7, 33.5 
emission in the datacube, nearly all of this component consisted of unrelated foreground emission (likely from the H~II region 
along the line of sight) as well as photo-ionized ISM surrounding \g292.  None of the ejecta in our IRS maps showed significant [S~III] emission,
although the partially radiative CSM shocks in the equatorial belt of \g292\, show weak [S~III] $\lambda$18.7
emission (Figure~6).  The presence of weak [S~III] from the equatorial belt is consistent with the presence of
partially radiative shocks detected in [O~III] in the Rutgers Fabry-Perot observations of \g292\, (Ghavamian, Hughes \& Williams
2005).  

The [Ne~III] and [O~IV] emission line maps are shown in Figure~4.  The pixel scales of the images are 5\arcsec pixel$^{-1}$, with
each image 11\farcm2$\times$11\farcm2 square.  They bear a strong resemblance to the [O~III] optical image
 (Winkler \& Long 2006), indicating that the oxygen and neon originate from the same nucleosynthetic layers within
the progenitor.   
The most prominent O-rich structure, the Spur, is clearly detected in both line maps, while the FMKs seen near the
northern and southern edges of \g292\, in the [O~III] images have faint counterparts in [Ne~III] and [O~IV].
The faint bands of [O~III] emission stretching southward from the Spur (the `Streamers') and westward toward the middle
of \g292\, are also detected in the IR.  However, there are also some differences between the [Ne~III] and [O~IV] images.
The shape of the Spur differs slightly in the two images, with the [O~IV] emission having a somewhat clumpier morphology
than the [Ne~III] emission.  In addition, the FMKs near the southern edge of \g292\, are more prominent in the [Ne~III]
images.  In particular, the southernmost FMK (an ejecta knot with one of the highest proper motions in \g292\, (Winkler \etal\,
2009) and which we name `Runaway FMK') is prominent in the [Ne~III] image, but barely detected in [O~IV].  This may be due in 
part to differences in the continuum subtraction between the two images.  Both the sky and SNR
continuum just begin to turn on near 15 $\micron$, and increase steadily past 26 $\micron$.  Subtraction of this
continuum adds more noise to the resulting [O~IV] image than to the [Ne~III] image in Figure~4, making it more difficult
to detect intrinsically faint features such as the FMKs in [O~IV].

Summing the emission over all the
radiatively shocked ejecta in \g292, we find that $L_{[Ne~III]}\,\approx\,$8.2$\times$10$^{33}$ \ergss\, and
L$_{([O~IV]+[Fe~II])}\,\approx\,$2.6$\times$10$^{34}$ \ergss, assuming a distance of 6 kpc.  The [O~IV]+[Fe~II] luminosity
of \g292\, (which should be close to the intrinsic value due to the minimal impact of interstellar reddening at 25 $\micron$) is 
approximately six times lower than the unreddened [O~III] luminosity (1.6$\times$10$^{35}$\,$d_6^2$ \ergss)
reported by Winkler \& Long (2006).
On the other hand, the [Si~II] $\lambda$34.8 emission
map shows no emission from the Spur. Save for the belt of O-rich material running westward from the Spur (not to be
mistaken with the circumstellar belt seen in X-rays), none of the radiatively shocked O-rich ejecta produce significant
[Si~II] emission.  The only feature clearly visible in the [Si~II] image in Figure~4 is a blob of emission located just
interior to the Spur.  Although the [Si~II] image is considerably noisier than the [Ne~III] and [O~IV] images and
suffers from fixed pattern residuals leftover from the hot pixel interpolation, it is evident that there is no
substantial [Si~II] emission from the radiatively shocked, optically bright ejecta in \g292.

\subsubsection{15-25 $\micron$ Continuum Emission }

The ability to generate images of \g292\, in isolated spectral ranges from our IRS maps allows us to isolate regions
of pure continuum emission to search for emission from shock-heated ejecta dust.
In our earlier spectroscopic study of \g292\, (Ghavamian \etal\, 2009) we identified a broad bump of emission
arising from the lower section of the O-rich Spur and centered near 18 $\micron$ and extending from 15 $\micron$ to 25 $\micron$.
Aside from this prominent bump, the only other observed spectral features from the Spur were emission lines.  The
spectral bump is a possible signature of protosilicate dust emission from Mg$_2$SiO$_4$ or MgSiO$_3$, and has been detected
in \spitzer\, IRS maps of 1E0102$-$72.3 (Sandstrom \etal\, 2009; Rho \etal\, 2009) and Cas A (Rho \etal\, 2008; Smith
\etal\, 2009).  Detection of the 15-25 $\micron$ bump in \g292\, could be evidence that dust formed in the SN ejecta,
and that this dust is currently being heated in the radiative ejecta shocks.
However, given the complicated mixture of the shocked and photoionized CSM (as well as unrelated foreground emission from the nearby H~II region) 
overlying the Spur in \g292, we sought to confirm the spatial coincidence of the emission bump with the supernova ejecta.

To this end, we used CUBISM to subtract the line of sight background from the datacube using emission off the
eastern edge of \g292\, (we used sky in this region because it is closest to the Spur).  We then integrated
the emission between 15 $\micron$ and 25 $\micron$ from our IRS LL2 datacube of
\g292, while excluding the [Ne~III] $\lambda$15.5 line emission from the ejecta in this bandpass.  Although there is
no evidence of significant [S~III] $\lambda$18.7 emission from the ejecta, we excluded this emission
line as well during the integration.  This sky-subtracted continuum image of \g292\, in the 15-25 $\micron$ range is shown in the
lower right panel of Figure~4.

The 15-25 $\micron$ map shows extensive continuum emission from \g292, arising almost
entirely from shock-heated CSM dust.  The equatorial belt and southwestern regions (e.g., Regions 13-16 and 6-10)
are especially prominent, while
the lower density filaments along the elliptical shell (e.g., regions 1,2,11 and 12 from Figure~3) are fainter.  These trends
are consistent with the dust temperature variations reflected in the 70/24 flux ratio.  The continuum from most of the CSM shocks
begins near 15 $\micron$, then rises steadily past the red end of the IRS bandpass at 36 $\micron$.  Remarkably, however,
some of the radiatively shocked O-rich ejecta (such as the Spur) are also detected in the 15-25 $\micron$ map.  
The Runaway FMK (marked by a red arrow at the bottom of each panel in Figure~4), which
has been thrown clear of most of the shocked CSM in the
interior of \g292, can also be seen.  The knot is located in a region of lower, less complicated background emission, allowing the emission
from this knot to stand out more easily than for the other FMKs lying in the projected interior.  The detection of 15-25 $\micron$
IR continuum from both the Spur and the Runaway FMK is strong evidence that whether
the emission arises from a 15-25 $\micron$ `bump' or some other type of emission feature, dust grains most likely formed 
in the O-rich ejecta and are currently being heated by radiative shocks.  

\section{DISCUSSION}

\subsection{MIPS 70/24 ratios}

The effective temperatures of heated interstellar dust grains in SNRs are largely dependent on the
post shock gas density (Dwek 1987; Dwek, Foster \& Vancura 1996), and hence, by extension, on the preshock gas density.  
At the high gas temperatures encountered in young non-radiative SNRs, the
impact of electrons from the hot postshock plasma is the primary source of dust heating.  On the other hand, dust sputtering in these shocks 
is primarily caused by proton impacts.  At these high temperatures, the amount
of energy deposited per electron impact into the dust grains is approximately constant, so that the equilibrium dust temperature (and hence IR emissivity)
of the dust depends mainly on the electron (and hence gas) density (e.g., Figure~6b from Dwek 1987 and Figure~8 from Dwek, Foster \& Vancura 1996).  Assuming
similar gas temperatures, compositions and column densities in the CSM shocks (consistent with the X-ray spectral fits of
Lee \etal\, (2010)), regions with bright X-ray emission should broadly correlate with regions of high gas density, which in turn correlate with
regions having bright IR emission and high dust temperatures.
The brightest X-ray emission should then correspond to IR emission with the smallest 70/24 
flux ratios. 

Comparing the MIPS and \chan\, X-ray images of \g292, we find that there does appear to be a correlation between
the 24 $\micron$ and 70 $\micron$ surface brightnesses and the X-ray surface brightness for most of the CSM shocks in \g292.
Interestingly, however, there are departures from this relationship observed along the southwestern side of \g292.
Specifically, the surface brightnesses of the southwestern CSM clumps (corresponding to Regions 6-10 in Figure~3) are 
$\sim$3 times higher than the rest of the shell in both IR bands, yet their X-ray counterparts are either faint or almost
non-existent.  The lack of bright X-ray emission matching the bright IR clumps in the SW is not due to enhanced local
absorption of X-rays, since Lee \etal\, (2010) did not find a large enough variation in column density 
(5$\times$10$^{21}\,\lesssim\,N_H\,\lesssim$ 7$\times$10$^{21}$) around the rim to account for the reduced
X-ray flux in the SW corner of \g292.

To quantify the relationships described above, we plotted the 24 $\micron$ and 70 $\micron$ surface brightnesses of
the selected regions in \g292\, against the X-ray surface brightnesses of those regions in Figure~5 (top panel).  The 70/24 ratios
(which are plotted against the X-ray surface brightnesses in the lower panel of Figure~5) show a noticeable declining tend with
increasing X-ray emission, consistent with our prediction.
One caveat to consider when interpreting Figure~5 is that X-ray emission from the ejecta overlies the emission from some of the CSM shocks.
Therefore, the X-ray count rates in some of the CSM regions plotted in Figure~5 are over-estimated to various degrees (save
for Region 4, most of the regions contain little or no discernible emission from ejecta).
The ejecta contribution to the region counts is likely to add scatter to the values along the horizontal axis of the plot.  However, the plot is 
still useful for identifying systematic trends between the IR and X-ray properties of \g292, and they clearly demonstrate a relationship
between emission in the two bands.  The 70 $\micron$ fluxes show more scatter than the 24 $\micron$ fluxes and their distribution
appears somewhat flatter than the 24 $\micron$ when plotted versus X-ray surface brightness.  Dividing the two fluxes results in
a more distinct correlation, with the 70/24 ratio declining with X-ray surface brightness.

The trend between the 70/24 ratio and the X-ray surface brightness is also consistent with predictions from the shock models of
Dwek, Foster \& Vancura (1996).  Specifically, the lowest ratios ($\sim$1.5-1.8) are
found in the denser material of the equatorial belt, while the largest ratios ($\sim$2.5-5) are found in the low density, fainter
sections of the outer shell.  A more quantitative comparison can be made with the Dwek, Foster \& Vancura (1996) predictions
by using dust spectra for their 800 \kms\, shock models (this shock speed gives a temperature closest to the average of the temperatures
of the different regions measured by Lee \etal\, (2010)).  From Figure~8c of Dwek, Foster \& Vancura  (1996), the dust spectra for preshock
gas densities of 0.1 cm$^{-3}$ and 1.0 cm$^{-3}$ should be most appropriate for modeling the IR emission from
the equatorial belt and circumstellar shell in \g292, respectively.  The dust spectra from Dwek, Foster \& Vancura (1996)
predict $\nu_{70}\,F(70)/\nu_{24}\,F(24)\,\approx\,$2.3 for n\,=\,0.1 cm$^{-3}$ and $\nu_{70}\,F(70)/\nu_{24}\,F(24)$ 
\,$\approx$\,0.7 for n\,=\,1 cm$^{-3}$.  The predicted flux ratios are then $F(70)/F(24)\,\approx\,$6.7 for n\,=\,0.1 cm$^{-3}$ and
$F(70)/F(24)\,\approx\,$2 for n\,=\,0.1 cm$^{-3}$.  These compare favorably with the observed ratios for the equatorial belt
and outer shell in Table~1, though the observed ratios for the shell are somewhat lower than predicted. This is likely
due to the inferred preshock density of the shell obtained by Lee \etal\, (2010) ($\sim$0.3-0.5 cm$^{-3}$) being slightly larger
than that of the n\,=\,0.1 cm$^{-3}$ model of Dwek, Foster \& Vancura (1996).

Overall, the regions with the hottest dust in Figure~5 are (not suprisingly) Regions 15, 17 and 18, corresponding to the brightest clumps
in the \chan\, image of \g292.  Park \etal\, (2002) found that the equatorial belt (Region 15) had a higher density and slightly lower temperature than 
the outer shell of \g292, while Ghavamian, Hughes \& Williams (2005) detected faint [O~III] $\lambda$5007 emission from
the belt.  This indicates that the elevated density in the equatorial belt in Region 15 has caused the shocks to become
partially radiative at that location.  The elliptical shell surrounding \g292, where the blast
wave propagates through the relic red supergiant (RSG) wind (Lee \etal\, 2010), has the lowest preshock gas densities, and hence the coldest
shocked CSM dust.  

Figure~5 provides an important clue to the origin of the elevated IR to X-ray flux ratios in the southwestern portion of 
\g292. Although the emission in the southwest is bright in both MIPS images, the 70/24 ratios from this part of \g292\,
(as reflected in the 70/24 ratios for the CSM knots in Regions 6-10) are similar to those of the equatorial
belt, where emission is bright in both IR and X-rays.   This suggests that the dust is just as hot in the southwestern
CSM knots as they are in the belt (as reflected in the 70/24 ratios for Regions 13-18).
This in turn implies that the gas densities in the southwest and in the equatorial belt are similar.  
Therefore, the most likely explanation the anomalously
high IR to X-ray ratios of Regions 6, 7, 9 and 10 is that these clumps have higher dust-to-gas ratios than the 
clumps in the equatorial belt.  A localized, elevated dust content in the southwest of \g292\, may indicate that the dust
in this portion of the SNR is not well coupled to the X-ray emitting gas, or it may reflect variations 
in the dust condensation efficiency in the relic red giant wind of the progenitor.  Even allowing for variations in dust
content around the rim of \g292, the average dust-to-gas ratio inferred by Lee \etal\, (2009) for this SNR ($\sim$10$^{-3}$)
is significantly lower than the average Galactic value (Weingartner \& Draine 2001).

In Figure~6 we show sample IRS mapping spectra extracted from four of the regions marked in Figure~3 (Regions 4, 7, 15 and 17).  
The spectra were obtained from a section of the blast wave in the NW (Region 4), a knot of bright IR emission in the SW
(Region 7), the brightest section of the equatorial belt (Region 15) and a large clump just above the eastern 
edge of the equatorial belt. We subtracted the same background emission from all four of the IRS spectra after averaging
the emission from two sky regions located just off the eastern and western sides of \g292\, (marked by the cyan boxes in the
left panel of Figure~3).  

The background-subtracted spectra from the four regions described above are dominated by a rising dust continuum.  
The blast wave emission from Region 4 shows a slowly rising continuum
which starts near 15 $\micron$ and rises past the red end of the IRS bandpass.  In contrast, the clump in
Region 7 shows a brighter, more steeply rising dust continuum which peaks at a shorter wavelength than
Region 4 (at approximately 33 $\micron$).  This is consistent (based on the discussion in Section 4.1) with a 
higher overall density in region 7 than Region 4.  Region 15, while showing a prominent dust continuum,
also exhibits faint emission lines of [Ne~III] $\lambda$15.5, [O~IV]+[Fe~II] $\lambda$25.9, [S~III] $\lambda$18.7
and [Si~II] $\lambda$34.8.  Region 15 has been detected in [O~III] in the optical (Ghavamian \etal\, 2005),
and was shown by Park \etal\, (2002) to be both denser and cooler than the blast wave filaments encircling \g292.  
The shocks in the equatorial belt have formed partial cooling zones and hence are partially radiative, giving
rise to the observed faint IR (and optical) line emission.  The continuum from Region 17 is similar to that of Region 15,
indicating similar dust temperature.  The presence of faint [Ne~III] $\lambda$15.5 and [O~IV]+[Fe~II] $\lambda$25.9
in the spectrum
of Region 17 indicates that the shocks in this clump have also started to form radiative cooling zones.

\subsection{The [Si~II] emission and the thermodynamic state of the ejecta}

In their analysis of the 510 ks \chan\, image of \g292, Park \etal\, (2007) concluded that the thermodynamic state
of the ejecta (as reflected by temperature, density and ionization state) exhibits a significant gradient between the
southeastern and northwestern sides of \g292.  The X-ray hardness ratios they obtained from the \chan\, data indicated
a substantially lower temperature for the SE ($k T\,\sim$0.7 keV) than the W and NW ($k T \sim$ 5 keV).  This is consistent
with the fact that most of the optical and IR-emitting SN ejecta -- the Spur and its associated streamers -- are found 
in the SE of \g292.  The ejecta shocks in the rest of the SNR, by contrast, are mostly still in the non-radiative phase.

The trends described above can be seen in the narrowband X-ray images of \g292\, in O He $\alpha$
and Ne He$\alpha$ (Park \etal\,2007).  Thse images show a strong spatial correlation with the optical [O~III] and IR [Ne~III] and [O~IV] emission in the SE.
In contrast, the narrowband X-ray images in Si He$\alpha$ (which traces hotter gas than O Ly$\alpha$ and Ne He $\alpha$ owing
to its higher nuclear charge) is almost entirely absent in the SE.  The lack of 
Si emission from the SE in both the X-rays and the IR (save for the isolated [Si~II] ejecta blob
in Figure~4) indicates that the reverse shock has not yet encountered most of the Si-rich ejecta in the SE.
Park \etal\, (2007) speculated that the cooler thermodynamic state of the ejecta in the SE of \g292\, was the result
of an asymmetric supernova explosion, where less energy was channeled into that direction than the rest of the SNR.  The lack of
extensive [Si~II] $\lambda$34.8 emission in the SE of \g292\, may indicate that the SNR is less evolved in
that direction, providing further evidence in favor of the asymmetric explosion picture.

\subsection{15-25 $\micron$ Bump}

As shown in Figure~4, the O-rich Spur and the Runaway FMK in \g292\, are detected in the 15-25 $\micron$
continuum image extracted from the \spitzer\, IRS datacube of \g292.  The presence of such continuum
indicates that dust grains exist within the ejecta.  To investigate this possibility further, we extracted IRS
spectra of the Ruanway FMK, using a circular region centered on the knot.  We subtracted sky emission
using an identically sized region located approximately 10\arcsec\, radially inward from the FMK (marked by the small cyan
circle in Figure~3).
The resulting spectrum is shown in Figure~7.   The sky spectrum shows emission features from intervening photodissociation region (PDR) and
H~II regions along the line of sight to \g292: PAH emission at 16.4 and 17.0 $\micron$, as well as H$_2$ S(1) 17.1 $\micron$ and 
H$_2$ 28.2 $\micron$ emission.   Line of sight emission from [Ne~III] $\lambda$15.5, [S~II] $\lambda\lambda$18.7, 33.5,
[Fe~II]+[O~IV] $\lambda$25.9 and [Si~II] $\lambda$34.8 are also present, as is continuum emission from
photo-heated dust.  After subtraction of the background, both [Ne~III] and [Fe~II]+[O~IV] emission remain,
signatures of the O- and Ne-rich ejecta.  However, the 15-25 $\micron$ bump also remains.  The spectral
characteristics of this feature are very similar to those of the Spur (Ghavamian \etal\, 2009), and similar to what is seen
in the radiatively shocked O-rich ejecta of 1E0102$-$72.3 (Sandstrom \etal\, 2009; Rho \etal\, 2009).  This strengthens the
case for SN dust (Mg$_2$SiO$_4$/MgSiO$_3$) in the ejecta of \g292.  It also dispels the possibility raised by
Lee \etal\, (2009) that oversubtraction of the overlying dust continuum was responsible for the lack of significant 
24-36 $\micron$ continuum emission in our earlier IRS
spectrum of the Spur (Ghavamian \etal\, 2009).  

\subsection{The Narrow Tail}

The elongated structure extending southward of \g292\, in the 24 $\micron$ and 70 $\micron$
images exhibits a very interesting morphology.  It was first noted by Lee \etal\, (2009) in their AKARI
observations of \g292, who referred to feature as a `Narrow Tail'.  It appears to be connected to
the streamers of O-rich ejecta
extending south of the Spur (seen in the [O~III] image in Figure~1, and in the 24 $\micron$
MIPS image).  However, the Narrow Tail has no X-ray counterpart, and no corresponding optical emission.  The 
70/24 ratio (Region 19 in Table~1) is 4.2, which indicates cold dust.  The background-subtracted
IRS spectrum of the Narrow Tail is shown in the right panel of Figure~7.  The location used for the
estimating the background contribution to the Narrow Tail spectrum is marked by the elongated cyan region in Figure~3.
The spectrum of the Narrow Tail in Figure~7 is consistent with cold dust -- it peaks beyond
the IRS bandpass.   These colors
are similar to IR cirrus, rather than dust heated by UV from the core collapse explosion.
Dwek \& Arendt (2008) presented an analysis of light echoes near Cas A, and found that
the 70/24 ratios of these echoes was significantly smaller than seen in normal
IR cirrus, an indication that the dust had been significantly heated.  Dwek \etal\, found
that graphite dust was required to explain the emission, finding that the dust temperature
($\sim$175 K) indicated very strong heating from incident UV radiation, rather than
optical photons from the SN.  In the case of \g292, the colors of the elongated structure
indicate much colder dust ($\lesssim$25~K), hence heating by UV emission from the
SN that produced \g292\, is highly unlikely.  There is weak [Si~II] $\lambda$34.8 
emission from the sky-subtracted spectrum of the Narrow Tail, as well as faint
[Fe~II]+[O~IV] and a hint of $H_2$ S(0,0) at 28.2 $\micron$.  
This is all consistent with a clump of cold interstellar cloud material (IR cirrus) which
has been heated by ambient stellar UV light.  We conclude that despite the very
suggestive morphology and alignment of this feature with the streamers of O-rich material
emanating from the Spur, it appears to be a chance alignment between an interstellar cloud
and the elongation axis of the O-rich streamers.

\section{SUMMARY}

We have presented MIPS 24 $\micron$ and 70 $\micron$ imaging and IRS spectral mapping of \g292\, obtained with \spitzer.
These observations complement the existing AKARI study of \g292\, (Lee \etal\, 2009).  Our results are as follows:

1.  The MIPS data show that most of
the filaments seen in the X-rays along the periphery of the SNR, as well as the band of equatorial material stretching
across its middle, emit dust continuum emission in the mid-IR.  The IRS mapping data of the filaments and clumps show broad-band, rising continua
between 15 $\micron$ and 40 $\micron$, a clear signature of shock-heated dust.  The shapes of the dust continua
are consistent with a mixture of graphite and silicate dust, as is observed in the non-radiative blast waves of other
SNRs (e.g, Williams \etal\, 2011).  

2. The MIPS 70/24 flux ratio varies significantly (1.5 $\lesssim$ $F_{70}/F_{24}$ $\lesssim$ 5) 
between the blast wave, equatorial belt and
the southwestern clumps.  These variations primarily reflect differences in dust temperature
around \g292, and plots of the 70/24 ratio versus X-ray surface brightness are consistent with variations 
in CSM density inferred from X-ray observations (Lee \etal\, 2010).  The 70/24 $\micron$ ratios are also
consistent with predictions from dusty shock models (Dwek, Foster \& Vancura 1996) with shock speeds
and preshock densities matching those 
predicted by the X-ray observations of Lee \etal\, (2010).

3. No mid-IR emission (either lines or continuum) is detected from the non-radiative,
X-ray emitting ejecta seen in the \chan\, images \g292.  The radiatively shocked O-rich ejecta are detected
in the MIPS 24 $\micron$ image, with most of the ejecta emission in that band consisting of [O~IV]+[Fe~II] 
$\lambda$25.9 emission.  Using an isolated FMK located near the southern edge of \g292\, we have confirmed 
the detection of the 15-25 $\micron$ emission bump from
the radiatively shocked O-rich ejecta (possibly Mg$_2$SiO$_4$ or MgSiO$_3$ dust formed in the ejecta).

4. Continuum-subtracted emission line maps of the [Ne~III] $\lambda$15.5 and [O~IV]+[Fe~II] $\lambda$25.9 emission
show a strong spatial correlation between the two, and emission in these lines is detected from both the Spur and the 
FMKs.  However, neither of these two structures is detected in the continuum-subtracted [Si~II] $\lambda$34.8 maps
of \g292.  Save for a localized blob of [Si~II] emission near the inner edge of the Spur, no obvious emission from
this species is seen elsewhere in the SNR.  Since the entire southeastern portion of \g292\, is also deficient in Si line
emission in the X-rays (Park \etal\, 2007), the lack of both IR and X-ray emission from Si in the Spur indicates
that most of the Si is unshocked in the southeastern quadrant of \g292 and that save for the [Si~II]-emitting blob, the reverse shock has not yet penetrated
most of the Si-rich ejecta in the southeast.  This 
may indicate that the southeastern portion of \g292\, is less evolved than the rest of the SNR, and may be further evidence 
of the asymmetric explosion scenario proposed by Park \etal\, (2007).

This work is based on observations made with the {\it Spitzer Space Telescope}, which is operated
by the Jet Propulsion Laboratory, California Institute of Technology under a contract with NASA.  Support
for the work of P. G. was supported by NASA through the \spitzer\, Guest Observer Program, as well as 
HST Grant GO-10916.08.  P. F. W. acknowledges support from the NSF through Grant AST-0908566.

\clearpage

\begin{deluxetable}{cccr}
\tablecaption{Surface Brightnesses of Features in MIPS 24 $\micron$ and 70 $\micron$ images of G292.0+1.8   }
\label{tblobs}
\tablewidth{0pt}
\tablehead{
\colhead{Region Number} & \colhead{F(24) (MJy sr$^{-1}$)}  &  \colhead{F(70) (MJy sr$^{-1}$)} & \colhead{$\frac{F(70)}{F(24)}$} \\
}
\startdata
1       &       0.74 $\pm$ 0.36 &       3.79 $\pm$ 0.83 &       5.14 $\pm$ 2.73 \\
2       &       1.68 $\pm$ 0.25 &       6.68 $\pm$ 0.57 &       3.98 $\pm$ 0.68\\
3       &       3.91 $\pm$ 0.36 &       10.4 $\pm$ 0.84 &       2.65 $\pm$ 0.32\\
4       &       1.93 $\pm$ 0.28 &       6.77 $\pm$ 0.65 &       3.50 $\pm$ 0.61\\
5       &       3.03 $\pm$ 0.39 &       10.5 $\pm$ 0.91 &       3.48 $\pm$ 0.54\\
6       &       6.47 $\pm$ 0.84 &       14.4 $\pm$ 1.85 &       2.23 $\pm$ 0.41\\
7       &       8.40 $\pm$ 0.97 &       17.4 $\pm$ 2.15 &       2.07 $\pm$ 0.35\\
8       &       2.79 $\pm$ 0.59 &       8.20 $\pm$ 1.38 &       2.94 $\pm$ 0.79\\
9       &       6.13 $\pm$ 0.64 &       16.4 $\pm$ 1.87 &       2.68 $\pm$ 0.38\\
10      &       5.62 $\pm$ 0.64 &       14.2 $\pm$ 1.47 &       2.52 $\pm$ 0.39\\
11      &       2.35 $\pm$ 0.24 &       7.62 $\pm$ 0.57 &       3.24 $\pm$ 0.41\\
12      &       2.12 $\pm$ 0.38 &       6.91 $\pm$ 0.87 &       3.26 $\pm$ 0.72\\
13      &       5.05 $\pm$ 0.72 &       11.1 $\pm$ 1.62 &       2.20 $\pm$ 0.45\\
14      &       5.05 $\pm$ 0.64 &       9.24 $\pm$ 1.33 &       1.83 $\pm$ 0.35\\
15      &       7.58 $\pm$ 0.53 &       12.2 $\pm$ 1.20 &       1.61 $\pm$ 0.19\\
16      &       5.80 $\pm$ 0.34 &       10.7 $\pm$ 0.76 &       1.84 $\pm$ 0.17\\
17      &       6.38 $\pm$ 0.44 &       8.79 $\pm$ 0.96 &       1.38 $\pm$ 0.18\\
18      &       3.18 $\pm$ 0.54 &       4.88 $\pm$ 1.19 &       1.54 $\pm$ 0.46\\
19      &       0.39 $\pm$ 0.19 &       1.67 $\pm$ 0.44 &       4.24 $\pm$ 2.34\\

\enddata
\end{deluxetable}


\begin{figure}[ht] 
   \centering
  \includegraphics[width=5.in]{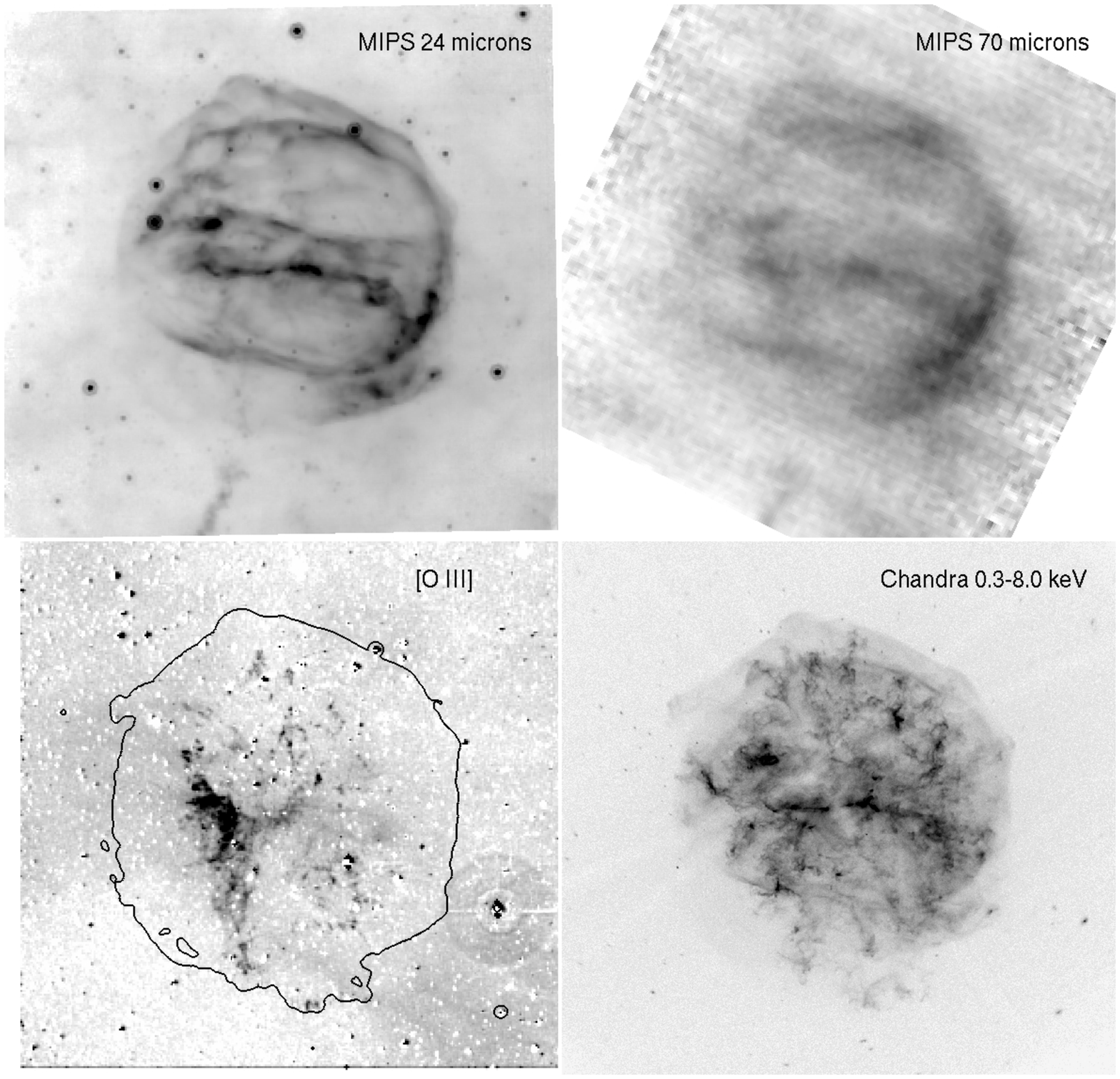}  
   \caption{MIPS images of \g292, shown alongside the optical [O~III] image (Winkler \& Long 2006) and the
510 ks \chan\, X-ray image (Park \etal\, 2007).   The outermost X-ray contour from the \chan\, image is marked
on the [O~III] image.  Each image is 12\farcm7$\times$13\farcm4 across, with East located at the left and
North at the top.
The emission in the 24 $\micron$ and
70 $\micron$ images is predominantly produced by shock-heated circumstellar dust.  The O-rich `Spur' is the bright, crescent-shaped
feature to the left of center in the [O~III] image.  The `Streamers' are the fainter fingers of emission stretching southward from
the Spur.  The fast-moving knots (FMKs) are the small knots of emission seen above and below the center of the [O~III] image.
A faint elongated feature
(named `the Narrow Tail' in Lee \etal\, 2009) in the south of \g292\, can be seen near the bottom of the two MIPS images.  The Narrow Tail
has no counterpart in the [O~III] or X-ray images.  
   }
   \label{fig:fig1}
\end{figure}

\begin{figure}[ht] 
   \centering
 \includegraphics[angle=0,width=6in]{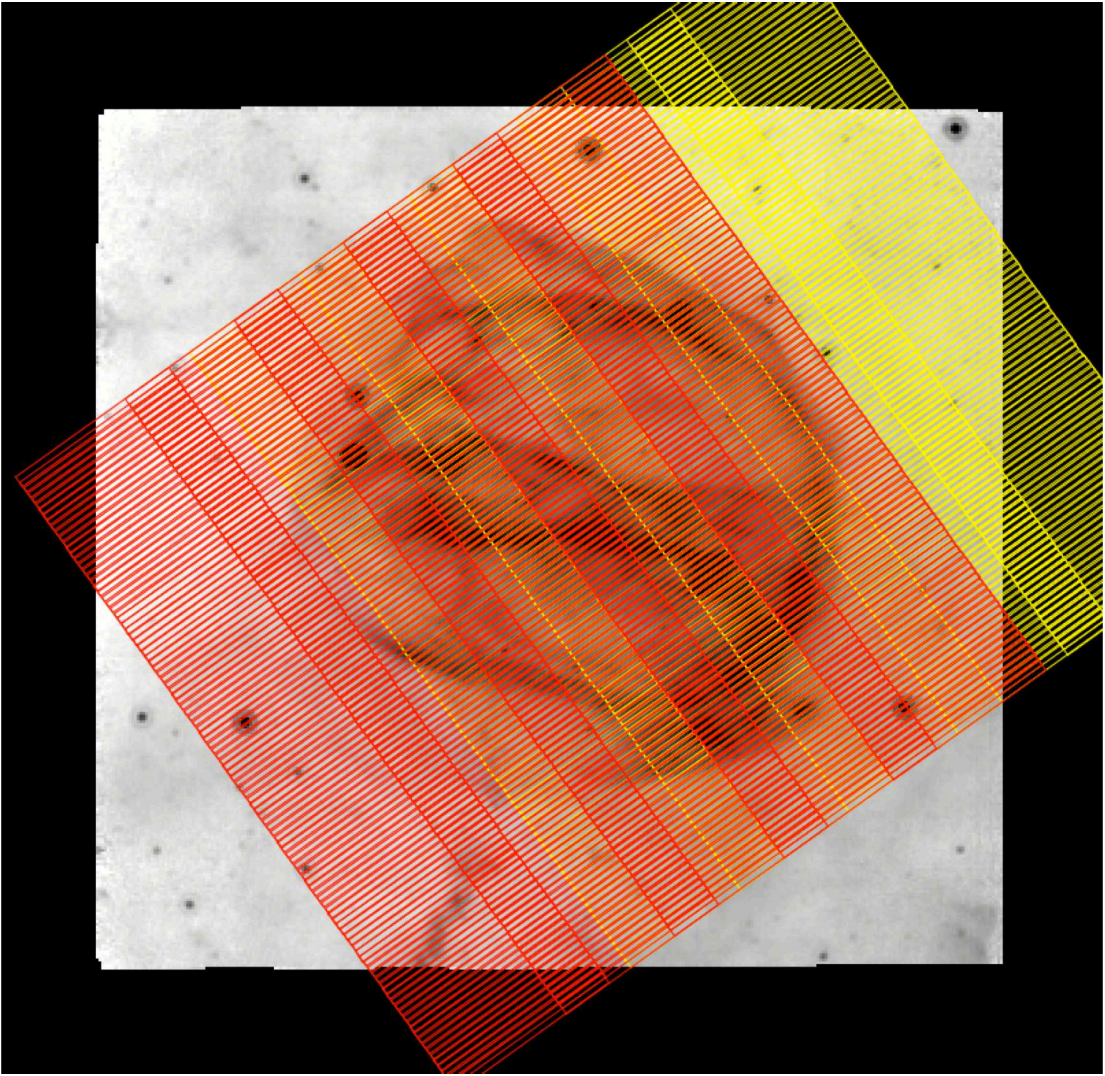} 
   \caption{
The IRS spectral mapping footprint, overlaid onto the MIPS 24 $\micron$ image of \g292\, for the
date of the IRS observations.  Long-Low 2 (14.2-21.8 $\micron$) slit positions are marked in yellow,
and Long-Low 1 (20.6-40 $\micron$) are marked in red. Slits are 10\farcs5\,$\times$\,168\arcsec\,
in size.
}
\label{fig:fig2}
\end{figure}

\begin{figure}[ht] 
   \centering
  \includegraphics[angle=270,width=7in]{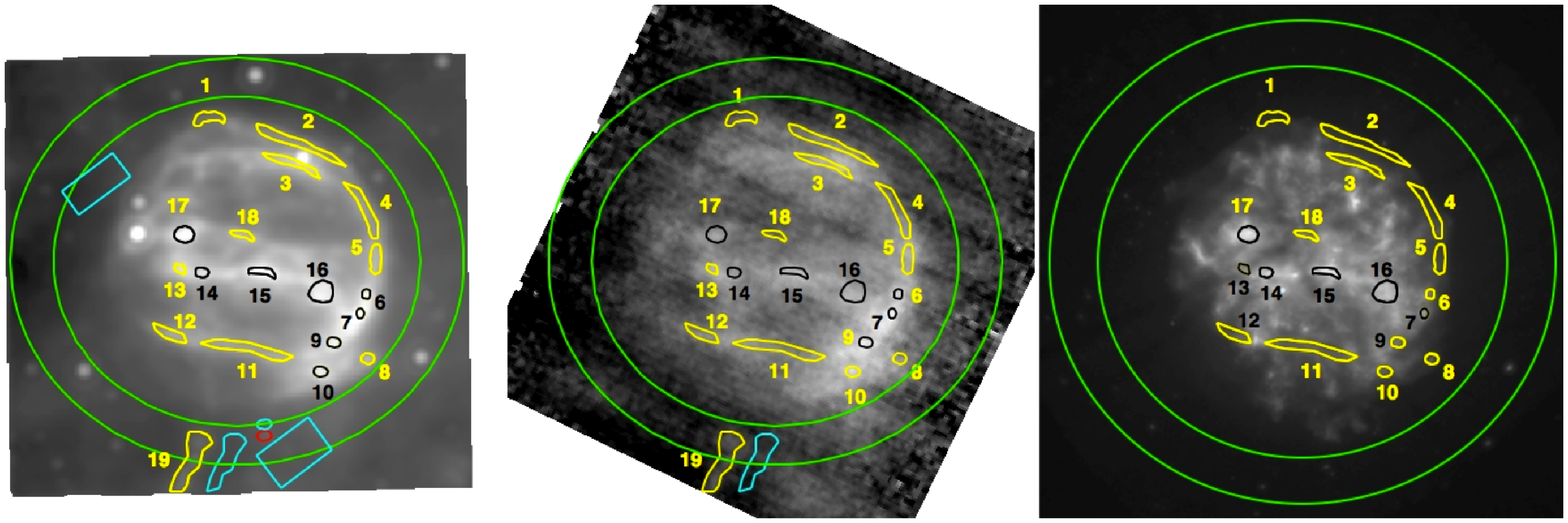}
   \caption{
  Left: The 24 $\micron$ MIPS image of \g292, shown with the flux extraction regions from Table~1 marked.
  The image has been convolved with a PSF kernel to match the 70 $\micron$ PSF of MIPS.  Middle: the
70 $\micron$ MIPS image with the flux extraction regions marked.    Right: The 510 ks \chan\, ACIS
image of \g292, blocked to 1\arcsec\, pixels and convolved with the MIPS 24 $\micron$ PSF, also with the 
Table~1 extraction regions marked.  This image is used
in computing the X-ray fluxes shown in Figure~5.   Region 19 (corresponding to the Narrow Tail) exhibits no
discernible X-ray emission, so is omitted in the X-ray flux measurements.  The red circular region on the 
24 $\micron$ image marks the extraction box for the Runaway FMK spectrum shown in Figure~7.  The green annuli marked around the SNR
are used for estimating and subtracting the background emission in each extraction region.  Although the background annuli
include a few point sources, these objects contribute negligibly to the summed background counts.  The background
used for generating the IRS spectra in Figure~6 was estimated by averaging the emission in the two cyan boxes
shown in the left panel.  The elongated and circular cyan regions were used for estimating the background emission
for the IRS spectra of the Narrow Tail and Runaway FMK, respectively, in Figure~7.  Some regions are colored in black
to create better contrast from the surrounding emission. }
\label{fig:fig3}
\end{figure}

\begin{figure}[ht] 
   \centering
 \includegraphics[angle=0,width=6in]{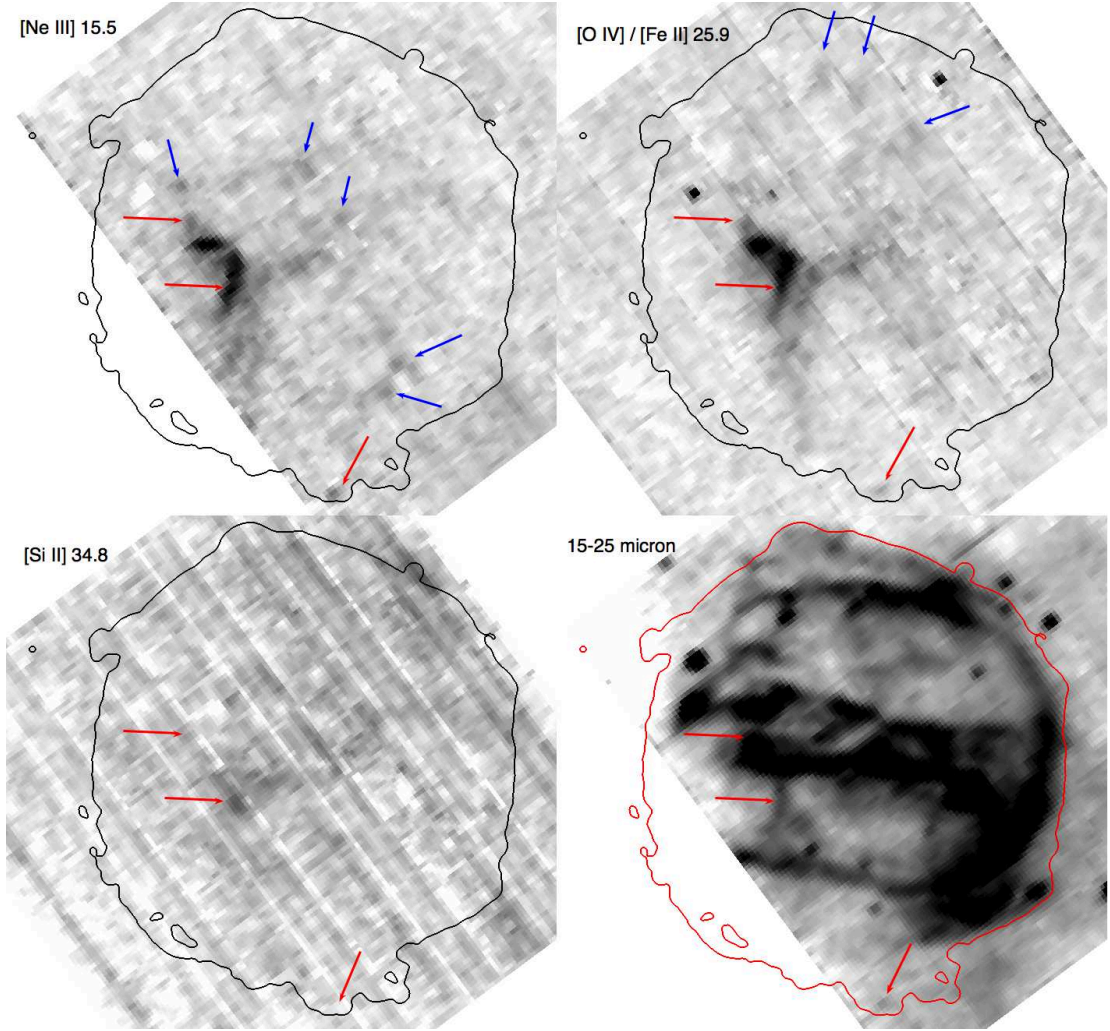} 
   \caption{Images of \g292\, in [Ne~III] $\lambda$15.5, [O~IV]+[Fe~II] $\lambda$25.9,
[Si~II] $\lambda$34.8 and 15-25 $\micron$ continuum, obtained from the IRS spectral map.   All images
have been background subtracted.  The contour in each image shows the outline of \g292\, 
in \chan\, 0.3-8.0 keV X-rays.  The red arrows indicate prominent regions where ejecta are detected
both in IR emission lines and in 15-25 $\micron$ continuum emission, with the latter revealing regions where
dust has formed in the SN ejecta.  The blue arrows mark the locations of prominent FMKs.  
  }
\label{fig:fig4}
\end{figure}

\begin{figure}[ht]
   \centering
 \includegraphics[angle=270,width=6in]{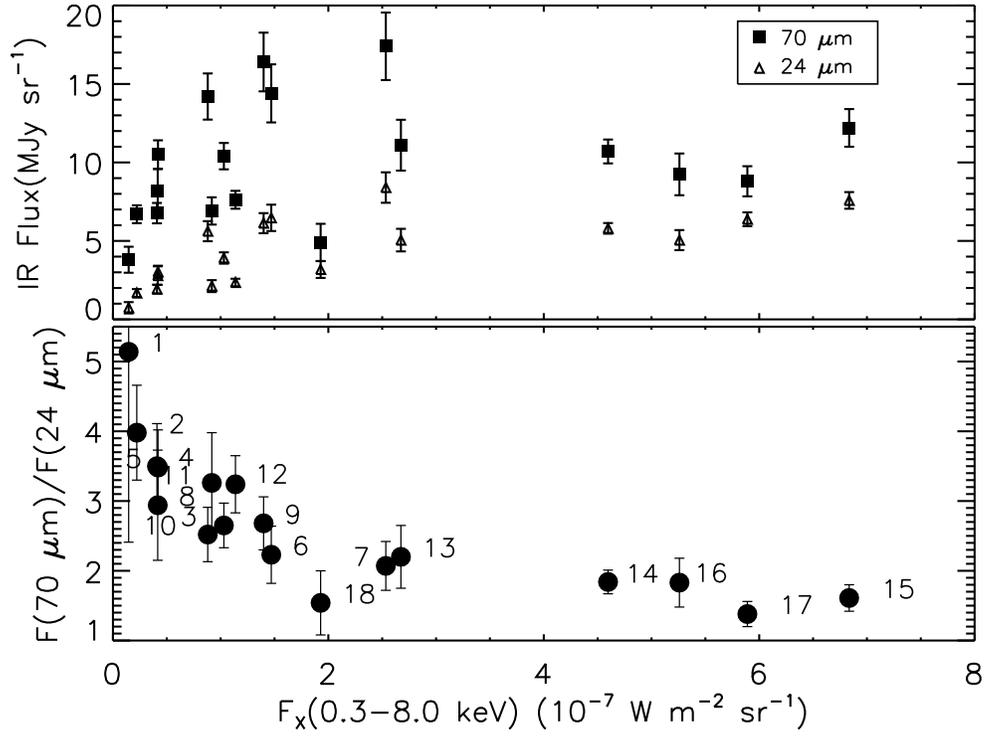}
   \caption{Top: the MIPS 24 $\micron$ and 70 $\micron$ surface brightness shown versus X-ray surface brightness for the flux extraction regions 
marked in Figure~3.  The 
X-ray surface brightness estimates assume shocked CSM, with a $kT\,=\,$0.75 keV APEC plasma model,
$N_H\,=\,$6$\times$10$^{21}$ cm$^{-2}$ and abundances 0.2 times solar.  The narrow tail (Region 19) exhibits
no discernible X-ray emission and is omitted from the plot.  Bottom: the 70/24 flux ratios for the same regions, shown versus
X-ray surface brightness.   
  }
\label{fig:fig5}
\end{figure}

\begin{figure}[ht]
   \centering
 \includegraphics[angle=270,width=7in]{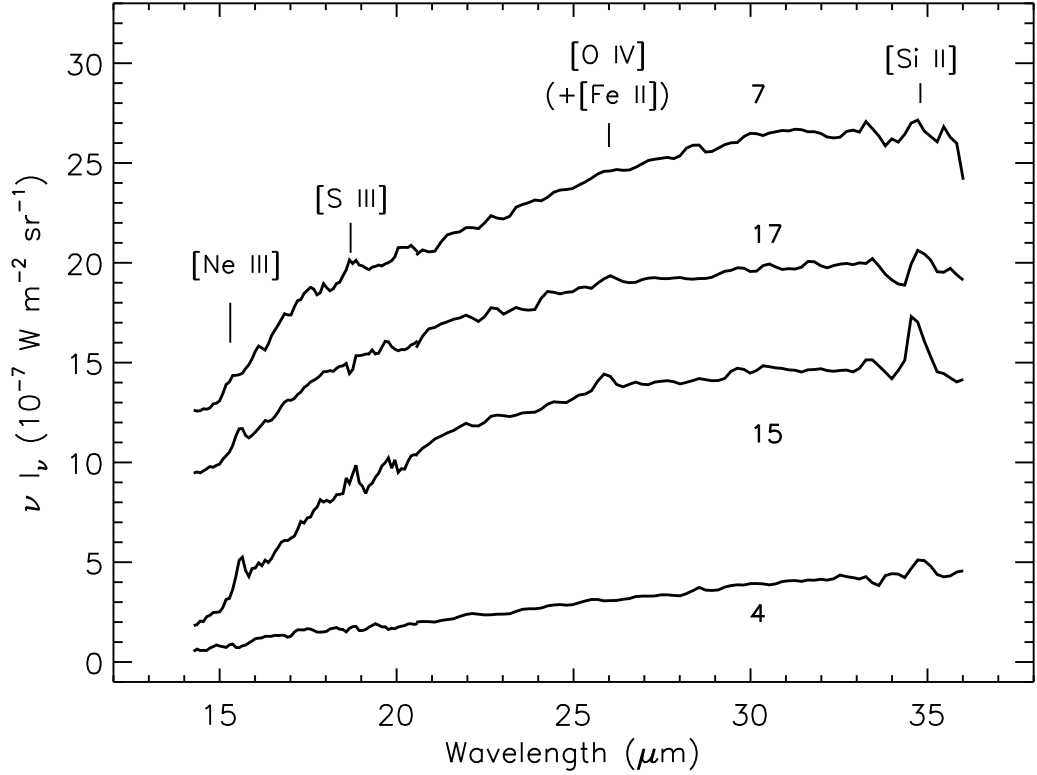}
   \caption{A sample of background-subtracted IRS mapping spectra of \g292, acquired from selected regions in Figure~3, as labeled.  
The subtracted background is the average of the emission on the eastern and western sides of the \g292\, (marked by the two cyan
boxes in the left-most panel of Figure~3).
The spectra have been shifted in flux to avoid overlap, with
Regions (4,7,15,17) being shifted by (0,10,0,7) units in $\nu\,I_{\nu}$.  The emission from each spectrum is dominated by shocked
CSM dust continuum, with the coldest dust found along the blast wave filaments (Region 4) and the hottest dust found in
the emission clump above the eastern edge of the equatorial belt (Region 17).  In addition to the CSM continuum in Region 15, the spectrum
at that location also shows faint emission from [Ne~III], [S~III], [O~IV]+[Fe~II] and [Si~II].  These lines are produced in slower,
partially radiative shocks in the equatorial belt (Park \etal\, 2002, 2004, Ghavamian \etal\, 2005).
  }
\label{fig:fig6}
\end{figure}

\begin{figure}[ht] 
   \centering
 \includegraphics[angle=270,width=7in]{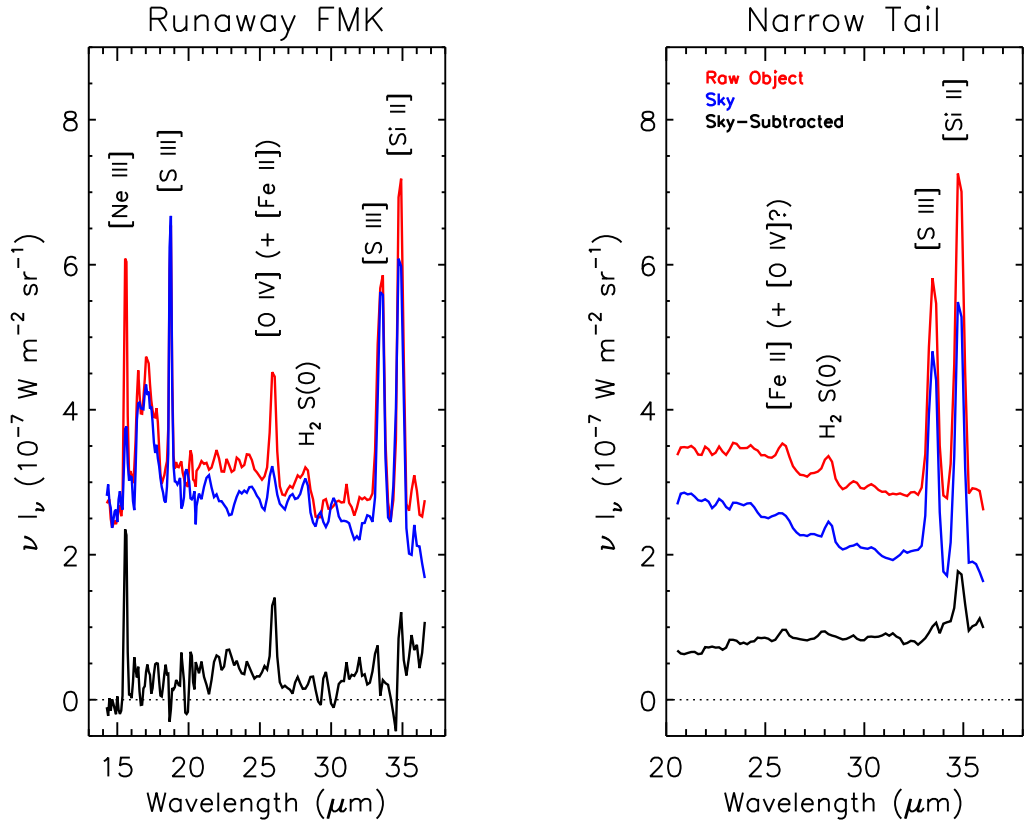} 
\caption{Extracted IRS mapping spectra of two features in \g292.  Left: combined LL2 and LL1 spectra of the southernmost FMK, showing the raw object
spectrum (red), the sky spectrum extracted from a nearby section of sky off the SNR (blue), and the resulting
sky-subtracted object spectrum (black).  The raw object and sky spectra have both been shifted by -2.0 units along the Y axis.
Right: The LL1 spectrum of the narrow tail of \g292, with similar color
schemes for the raw, sky and sky-subtracted object spectrum.  The sky spectrum shows emission features from intervening PDR and
H~II regions along the line of sight to \g292: PAH emission at 16.4 and 17.0 $\micron$, as well as H$_2$ S(1) 17.1 $\micron$ and
H$_2$ 28.2 $\micron$ emission.   The raw object and sky spectra have both been shifted by +2.0 units along the Y axis.
  }
\label{fig:fig7}
\end{figure}

\end{document}